\documentclass[12pt]{article}
\pdfoutput=1
\usepackage{epsfig, amssymb, graphicx, amsmath} 

\usepackage{empheq} 	
\usepackage{enumitem}
\usepackage{eurosym}	
\usepackage{wrapfig}	
\usepackage{float}		
\usepackage{subfig}	
\usepackage{footnote}
\usepackage{bm}         

\usepackage{bbm}		

\usepackage[authoryear]{natbib}
\bibpunct{(}{)}{,}{a}{}{,}

\newtheorem{theorem}{Theorem}
\newtheorem{proposition}[theorem]{Proposition}
\newtheorem{lemma}[theorem]{Lemma}

\newcommand{\E}{\mathbb{E}}

\newcommand{\be}{\begin{equation}}
\newcommand{\en}{\end{equation}}
\newcommand{\ben}{\begin{equation*}}
\newcommand{\enn}{\end{equation*}}
\newcommand{\bea}{\begin{eqnarray}}
\newcommand{\ena}{\end{eqnarray}}

\begin{document}
 
\newlength\tindent
\setlength{\tindent}{\parindent}
\setlength{\parindent}{0pt}
\renewcommand{\indent}{\hspace*{\tindent}}

\begin{savenotes}
\title{
\bf{ 
Model risk in mean-variance portfolio selection:
an analytic solution to the worst-case approach 
}}
\author{
Roberto Baviera$^\ddagger$ \& 
Giulia Bianchi$^\ddagger$ 
}

\maketitle

\vspace*{0.11truein}
\begin{tabular}{ll}
$(\ddagger)$ &  Politecnico di Milano, Department of Mathematics, 32 p.zza L. da Vinci, Milano \\
\end{tabular}
\end{savenotes}

\vspace*{0.11truein}

\begin{abstract}
\noindent
In this paper we consider the worst-case model risk approach described in \citet{Glasserman2014}. 
Portfolio selection with model risk  can be a challenging operational research problem. 
In particular, it presents an additional optimisation  compared to the classical one.
We find the analytical solution for the optimal mean-variance portfolio selection in the worst-case scenario approach and 
for the special case with the additional constraint of a constant mean vector considered in \citet{Glasserman2014}.

\noindent
Moreover, we prove in two relevant cases --the minimum-variance case and the symmetric case, i.e. when all assets have the same mean-- 
that the analytical solutions in the alternative model and in the nominal one are equal; we show that this corresponds to the situation when model risk reduces to estimation risk. 
\end{abstract}

\vspace*{0.11truein}
\vspace*{0.11truein}


\vspace{4cm}
\begin{flushleft}
{\bf Address for correspondence:}\\
Roberto Baviera\\
Department of Mathematics \\
Politecnico di Milano\\
32 p.zza Leonardo da Vinci \\ 
I-20133 Milano, Italy \\
Tel. +39-02-2399 4575\\
Fax. +39-02-2399 4621\\
roberto.baviera@polimi.it
\end{flushleft}

\newpage

\vspace*{0.21truein}


\section{Introduction}

\citet{Markowitz1952} was the first to introduce an optimal portfolio selection according to the mean and the variance. 
Since that seminal paper, this problem has been extensively studied \citep[see e.g.][and references therein]{LiNg}.
This criterion is at the base of modern portfolio theory and it is widely used in finance due to its simplicity given that
it models asset returns as Gaussian random variables.

The accuracy of this portfolio selection crucially depends on the reliability of this model, which is named nominal model. 
Model risk is the risk arising from using an insufficiently accurate model.
A quantitative approach to model risk is the worst-case approach, which was introduced in decision theory by \citet{GilboaSchmeidler1989}.
According to this methodology, one considers a class of alternative models and minimises the loss encountered in the worst-case scenario. 

The literature distinguishes between {\it estimation} and {\it misspecification} risk \citep[see e.g.][]{Kerkhof2010}.
In general, it is interesting to identify vulnerabilities to model error that result not only from
parameter perturbations ({\it estimation} risk)
but also from an error in the joint distribution of returns ({\it misspecification} risk).
The deviation between statistical distributions can be measured by the \citet{KullbackLeibler} relative entropy, which is also known as KL divergence, 
as proposed by \citet{HansenSargent2008} in the context of model risk.
The problem of determining an optimal robust portfolio under KL divergence has been studied by \citet{Calafiore2007}; he proposed two numerical schemes to find an optimal portfolio in the mean-variance and the mean-absolute deviation cases, considering a discrete setting.
This approach has been studied by \citet{Glasserman2014} in a continuous setting 
in a mean-variance case; 
the authors identified the worst-case alternative models
to the nominal model and numerically found the optimal portfolio selection in these cases.
More recently, \citet{Penev2019} have analyzed the mean-standard deviation case in detail showing that this case presents a semi-analytic solution.

\bigskip
Let us briefly summarise the portfolio selection problem in presence of model risk. 
Let ${\bf{X}} \in \mathbb{R}^n$ denote the stochastic asset returns. 
The p.d.f. associated with ${\bf{X}}$, $f({\bf{X}})$, corresponds to the nominal model, 
while the p.d.f. $\tilde{f}({\bf{X}})$ corresponds to the alternative model. 
The KL divergence between the two models is
\begin{equation}
\label{eqn:relentropy_introduction}
R(\tilde{f},f):=\mathbb{E}\left[m({\bf{X}})\ln{m({\bf{X}})}\right]
\end{equation}
where $m({\bf{X}}):=\tilde{f}({\bf{X}})/{f({\bf{X}})}$ is the change of measure and $\E[ \bullet ]$ denotes the expectation w.r.t. ${f}({\bf{X}})$.
In particular, we are interested in the alternative models
within a ball $ P_\eta$ of radius $\eta > 0$ around the nominal model; i.e., characterised by a KL divergence lower or equal to $\eta$.
 
Let $V_\textbf{a}({\bf{X}})$ denote 
a measure of risk associated with ${\bf{X}}$, 
that depends on the portfolio weights $\textbf{a}$ ranging over a set $\mathcal{A}$; 
the classical optimal portfolio selection problem is
\begin{equation}
\label{eqn:probopt1inf}
\inf_{\textbf{a}\in \mathcal{A}} \mathbb{E}[V_\textbf{a}({\bf{X}})],
\end{equation} 
while the worst-case portfolio selection corresponds to
\begin{equation}
\label{eqn:probopt1infsup}
 \inf_\textbf{a}\sup_{m\in P_\eta}\mathbb{E}[m({\bf{X}})V_\textbf{a}({\bf{X}})] \; .
\end{equation}

It can be shown that 
it is equivalent to the dual problem  \citep[see, e.g.][]{BoydConvexOptim}
\begin{equation}
\label{eqn:probLagrange_infinfsup}
\inf_\textbf{a}\inf_{\theta>0}\sup_m \mathcal{L}\left(\theta,\textbf{a}; m(\textbf{X}) \right)
\end{equation}
where 
\begin{equation*}
\mathcal{L}\left(\theta,\textbf{a}; m(\textbf{X})\right)=\mathbb{E}\left[m({\bf{X}})V_\textbf{a}({\bf{X}})-\frac{1}{\theta} \bigg( m({\bf{X}})\ln{m({\bf{X}})}-\eta \bigg) \right]
\end{equation*}
is the Lagrangian function associated to the constrained maximisation problem in \eqref{eqn:probopt1infsup}.

Thus, in the worst-case portfolio selection, one has to solve three nested optimisation problems where the inner problem is an infinite dimensional optimisation. 
While the inner optimisation problem is a standard one in functional analysis and a closed form solution can be found
\citep[see e.g.][]{Lam2016}, 
the presence of the other two makes the optimal selection a challenging operational research problem.
\citet{Glasserman2014} propose a numerical approach to solve this problem. 
In this study, 
we provide an analytical solution and
we show that the problem can be challenging from a numerical point of view.

\bigskip

This paper makes three main contributions.
First, we analytically solve the model risk optimisation problem in the worst-case approach when asset returns are Gaussian.
This result is achieved for a class of problems that are even wider than those solved numerically by \citet{Glasserman2014}. In particular, 
we consider 
\begin{itemize}
\item a generic mean-variance selection, and not just the case where we impose the additional constraint of the worst-case mean equal to the nominal one \citep[cf.][p.36]{Glasserman2014};
\item all possible values of $\theta$, which allow a  well-posed problem  and  we do not limit the analysis to 
``$\theta >0$ sufficiently small" \citep[cf.][p.31]{Glasserman2014}; i.e., we do not consider only small balls $P_\eta$.
\end{itemize}

Second, we provide the solution also 
in the special case where we impose the additional constraint of constant mean in the alternative model: 
this is the optimization problem considered by \citet[][cf. Eq.(30), p.36]{Glasserman2014}.

Third, we prove that,  in the minimum-variance case and
in the symmetric case with equal mean values for all assets in the portfolio,
the optimal worst-case portfolio is the same as the optimal nominal portfolio.
Moreover, we prove that in these cases model risk and \textit{estimation} risk coincide: 
we show that any alternative model within the ball $P_\eta$ can be obtained through a parameter change.
This result is different from the numerical solution in \citet[][Figure 1, p.37]{Glasserman2014}. 

\bigskip

The rest of this paper is structured as follows. In Section {\bf 2}, we recall the problem formulation. 
In Section {\bf 3}, we present model risk analytical solution in the mean-variance framework. 
In Section {\bf 4}, we study in detail the case of mean-variance with fixed mean considered by \citet{Glasserman2014}.
In Section {\bf 5}, 
we focus on the case where the optimal portfolio in the alternative model and the one in the nominal model coincide and  provide numerical examples.
Section {\bf 6} concludes this paper.

\section{Problem formulation}

In this section we recall the worst-case approach for model risk. 
Let ${\bf{X}}$ denote the stochastic element of a model and $\textbf{a}$ the parameters' vector ranging over the set $\mathcal{A}$; 
the nominal model corresponds to solve the optimisation problem \eqref{eqn:probopt1inf} in the nominal measure,
%
%
while the alternative model corresponds to the same problem with respect to an alternative measure, chosen within a KL-ball $P_\eta$ with $R(\tilde{f},f) < \eta$; i.e., within all models with a KL-divergence from the nominal model lower than a positive constant $\eta$.
In the best-case and in the worst-case approaches, the optimisation problem becomes
\begin{equation*}
\left\{
\begin{array}{ll}
\displaystyle \inf_\textbf{a}\inf_{m\in P_\eta}\mathbb{E}[m({\bf{X}})V_\textbf{a}({\bf{X}})]  & \text{best-case} \, , \\[4mm]
\displaystyle \inf_\textbf{a}\sup_{m\in P_\eta}\mathbb{E}[m({\bf{X}})V_\textbf{a}({\bf{X}})] & \text{worst-case} \, .
\end{array}
\right.
\end{equation*}
In portfolio selection, to have a robust measure, we are more interested in the worst-case approach, so hereafter we focus,
unless stated differently, on this case
that corresponds to the highest possible value of the measure of risk; 
{\it mutatis mutandis} similar results hold in the other case.

\bigskip

The rest of this section is organised as follows.
First, to clarify the notation used in the case of interest, we summarise the classical mean-variance 
portfolio theory with its main results \citep{Markowitz1952, Merton1972}. 
Then, we sum up the main results for the worst-case model risk approach in a rather general setting, following \citet{Glasserman2014} notation.


\subsection{Classical portfolio theory}

In this study, the nominal model is characterised by $n$ risky securities that are modeled as
a vector of asset returns $\textbf{X} \in \mathbb{R}^n$ distributed as a multivariate normal $\textbf{X}\sim N(\bm{\mu},\Sigma)$, with $\Sigma\in \mathbb{R}^{n \times n}$ 
a positive definite matrix with strictly positive diagonal elements.
Let ${\bf{a}}$ be the vector of portfolio weights, defined in the set $\mathcal{A}:=\{\textbf{a}:\, \textbf{a}^T\textbf{1}=1\}$, where $\textbf{1}$ is the vector in $\mathbb{R}^n$ of all $1$s.

\bigskip

In the mean-variance framework, one considers a quadratic measure of risk; 
i.e., the difference between the variance (multiplied by $\gamma$, a positive risk aversion parameter) and the expected return of the portfolio
\begin{equation}
\label{eqn:VaX}
V_\textbf{a}(\textbf{X}):=\frac{\gamma}{2} \, \textbf{a}^T(\textbf{X}-\bm{\mu})(\textbf{X}-\bm{\mu})^T \textbf{a}-\textbf{a}^T\textbf{X}, \;\;  \gamma > 0  \; .
\end{equation}

The 
value of the risk measure is 
\begin{equation*}
\mathbb{E}[V_\textbf{a}(\textbf{X})]=\frac{\gamma}{2} \, \textbf{a}^T\Sigma \, \textbf{a}-\textbf{a}^T\bm{\mu} \; .
\end{equation*}
The problem consists in minimising the value of the risk measure on all portfolios $\textbf{a}$ with weights summing to 1. Using a Lagrange multiplier, the mean-variance portfolio selection problem can be written as 
\begin{equation}
\label{eqn:frontierlagrange}
\min_\textbf{a} \left\{ \frac{\gamma}{2} \, \textbf{a}^T\Sigma \, \textbf{a} - \textbf{a}^T\bm{\mu} + \alpha \left(1-\textbf{a}^T\textbf{1}\right) \right\}
\end{equation}
where $\alpha$ is the multiplier.

Following \citet{Merton1972}, we introduce the notation 
\begin{equation}
\label{eqn:ABCD}
A:=\textbf{1}^T\Sigma^{-1}\bm{\mu} \qquad B:=\bm{\mu}^T\Sigma^{-1}\bm{\mu} \qquad C:=\textbf{1}^T\Sigma^{-1}\textbf{1} \qquad D:=BC-A^2 \; ;
\end{equation}
it is straightforward to show that $B, C > 0$ and $D \geq 0$ \citep[see, e.g.][]{Merton1972}.

The optimal mean-variance portfolio 
\citep[see, e.g.][equation (9), p.1854]{Merton1972} is
\begin{equation}
\label{eqn:astarclassic}
\textbf{a}^\star_{\text{nom}}=\displaystyle \frac{A}{\gamma}\;\frac{\Sigma^{-1}\bm{\mu}}{A}+ 
\left( 1- \frac{A}{\gamma} \right) \;\frac{\Sigma^{-1}\textbf{1}}{C} \; . 
\end{equation}

Any optimal portfolio $\textbf{a}^\star_{\text{nom}}$ is the linear combination of two portfolios in the optimal frontier  
$\textbf{a}_1^\star:={\Sigma^{-1}\bm{\mu}}/{A}$ and $\textbf{a}_0^\star:={\Sigma^{-1}\textbf{1}}/{C}$, where the latter is the portfolio of minimum variance.
This important result is also known as the two mutual fund theorem.

\subsection{Worst case model risk}
We briefly recall the model risk formulation for the construction of the alternative model. 
In particular, we focus on 
the worst-case portfolio selection  \eqref{eqn:probopt1infsup}; 
i.e., the one that considers the maximum value of the risk measure within the KL-ball $P_\eta$.
This worst-case problem 
is equivalent to the dual problem \eqref{eqn:probLagrange_infinfsup} 
with $V_\textbf{a}(\textbf{X})$ defined in (\ref{eqn:VaX});  {\it mutatis mutandis} the same result holds in the best-case with $\theta<0$. 

\bigskip 

{\it Remark}.
\citet{Glasserman2014} consider the special case with the additional constraint $\bm{\mu}=\mathbb{E}[m(\textbf{X})\textbf{X}]$; in their case, it is equivalent to consider, instead of \eqref{eqn:VaX}, the measure of risk
\begin{equation}
\label{eqn:VaGX_X}
V_{\textbf{a}}^{GX}(\textbf{X}) := \frac{\gamma}{2} \, \textbf{a}^T(\textbf{X}-\bm{\mu})(\textbf{X}-\bm{\mu})^T \textbf{a}-\textbf{a}^T\bm{\mu} \,, \, \gamma>0 \; .
\end{equation}
In Section 4, we show that all results obtained in the mean-variance framework hold even in this special case. 

\bigskip 

Thus, we have to consider 
the three nested optimisation problems in \eqref{eqn:probLagrange_infinfsup}. The inner optimisation problem is standard in functional analysis.
For a given $\theta>0$ and for a given $\textbf{a}\in \mathcal{A}$, the solution of the internal maximisation problem on the variable $m({\bf{X}})$ in \eqref{eqn:probLagrange_infinfsup} is
\begin{equation}
\label{eqn:mstarthetaa}
m^\star_{\theta,\textbf{a}}(\textbf{X})=\dfrac{\exp\left(\theta V_\textbf{a}(\textbf{X})\right)}{\mathbb{E}\left[\exp\left(\theta V_\textbf{a}(\textbf{X})\right)\right]} \; .
\end{equation}
This result is known in the literature \citep[see e.g.][]{Glasserman2014, HansenSargent2008}. 
For a complete proof, the interested reader can refer to \citet[][proposition 3.1]{Lam2016}. 
Unfortunately, the other two optimisations are more challenging and closed form solutions cannot be found in the literature for the case of interest.

\bigskip

Before entering into the details of the two optimisations in $\textbf{a}$ and $\theta$, it is interesting to observe some properties for the entropy computed on the optimal solution of the internal maximisation problem
\begin{equation}
\label{eqn:relentr_mstar}
R(\theta,\textbf{a}):=R(f \; m^{\star}_{\theta,\textbf{a}},f)
\end{equation}
where $R(\tilde{f},f)$ is defined in \eqref{eqn:relentropy_introduction}. 
They are stated in the following lemma. 
\begin{lemma}
\label{lem:relentr_monotone}
For any $(\theta, \textbf{a})$ s.t. $m^\star_{\theta,\textbf{a}}(\textbf{X})$ in \eqref{eqn:mstarthetaa} is well-defined,
$R(\theta,\textbf{a})$ is a monotone increasing function in $\theta>0$ for any portfolio  $\textbf{a}$
(and  monotone decreasing  for $\theta <0$).
\end{lemma}

{\bf Proof.}
See Appendix A    $ \qquad\hspace*{\fill}  \clubsuit \; $

\bigskip

Let us underline that the previous lemma shows a
general property that does not depend on the distribution of $\textbf{X}$ and on the measure of risk $V_\textbf{a}(\textbf{X})$.
As already stated in the introduction, in this paper we consider $\textbf{X}$ distributed as a multivariate normal $\textbf{X}\sim N(\bm{\mu},\Sigma)$ and the general mean-variance framework (i.e. with the measure of risk defined as in \eqref{eqn:VaX}). We now deduce a necessary and sufficient condition for which the change of measure \eqref{eqn:mstarthetaa} is well-defined and we find the distribution of $\textbf{X}$ in the alternative model for any portfolio $\textbf{a}\in \mathcal{A}$.

\begin{lemma}
\label{lem:ftilde}
Let $\textbf{X}\sim N(\bm{\mu},\Sigma)$, the change of measure $m^\star_{\theta,{\bf{a}}}(\textbf{X})$ in \eqref{eqn:mstarthetaa} 
is well-defined if and only if 
$\theta\in \left[0,\theta_{\text{max}}({\bf{a}})\right)$ where 
\begin{equation}
\label{eqn:conditiontheta}
\theta < \theta_{\text{max}}({\bf{a}}):=\dfrac{1}{\gamma \, {\bf{a}} ^T \Sigma\,{\bf{a}}} \; .
\end{equation}
Moreover,  for any ${\bf{a}}\in\mathcal{A}$,  in the alternative model $\tilde{f}(\textbf{X})$ corresponding to $m^\star_{\theta,{\bf{a}}}(\textbf{X})$,
$\textbf{X}$ is distributed as a multivariate normal r.v., i.e. $\textbf{X}\sim N(\bm{\tilde{\mu}},\tilde{\Sigma})$, where
\begin{equation}
\label{eqn:mutilde_Sigmatilde}
\bm{\tilde{\mu}}=\bm{\mu}-\theta\,\tilde{\Sigma}\,{\bf{a}} \qquad \tilde{\Sigma}=\left(\Sigma^{-1}-\theta\gamma{\bf{a}}\,{\bf{a}}^T\right)^{-1} \; .
\end{equation} 
\end{lemma}

{\bf Proof.}
First, 
we prove that $m^\star_{\theta,{\bf{a}}}(\textbf{X})$ in  \eqref{eqn:mstarthetaa} is well defined.
Let us observe that a necessary and sufficient condition to have a  well-defined
change of measure \eqref{eqn:mstarthetaa} is that $\mathbb{E}\left[\exp(\theta V_\textbf{a}({\bf{X}}))\right]$ is finite. 
We consider $\textbf{X}\sim N(\bm{\mu},\Sigma)$ and $V_{\textbf{a}}(\textbf{X})$ as in \eqref{eqn:VaX}, thus 
we get
\begin{equation*}
\mathbb{E}\left[\exp(\theta V_\textbf{a}({\bf{X}}))\right]=
\int d \textbf{X}
\dfrac{1}{\sqrt{(2 \pi)^n \det(\Sigma)}}
\exp\left(-\theta\,\textbf{a}^T\textbf{X}-\frac{1}{2} (\textbf{X}-\bm{\mu})^T\tilde{\Sigma}^{-1}\, (\textbf{X}-\bm{\mu})\right),
\end{equation*}
where $\tilde{\Sigma}^{-1}:=\Sigma^{-1}-\theta\gamma\,\textbf{a}\textbf{a}^T$ is a symmetric matrix.

The integral is finite if and only if the matrix $\tilde{\Sigma}^{-1}$ is positive-definite. 
To prove this fact we proceed in two steps: 
first we compute the determinant of $\tilde{\Sigma}^{-1}$ and we then state a property on the signs of its eigenvalues.

To compute the determinant, we use the Matrix Determinant Lemma \citep[see e.g.][theorem 18.1.1, p. 416]{HarvilleMatrix} that states 
\begin{equation}
\label{eqn:matrixdetlemma}
\det\left(\Sigma^{-1}-\theta\gamma\,\textbf{a}\textbf{a}^T\right)=\left(1-\theta\gamma\,\textbf{a}^T\Sigma\,\textbf{a}\right)\det\left(\Sigma^{-1}\right).
\end{equation}

Thus, 
the determinant of  $\tilde{\Sigma}^{-1}$ is positive if and only if the condition \eqref{eqn:conditiontheta} holds.
If $\det \tilde{\Sigma}^{-1}$ is positive, then $\tilde{\Sigma}^{-1}$ is also invertible; we define  $\tilde{\Sigma}$ as its inverse. 
We have verified that \eqref{eqn:conditiontheta} is a necessary  condition  to 
get $\tilde{\Sigma}^{-1}$, thus $\tilde{\Sigma}$, positive-definite.

\smallskip

We now prove that the condition $\theta<\theta_{\text{max}}(\textbf{a})$ is also
sufficient to have the matrix $\tilde{\Sigma}^{-1}$ positive-definite. 
Let $\lambda_i$ be the eigenvalues of $\Sigma$. The eigenvalues of the inverse matrix $\Sigma^{-1}$ are the reciprocals $1/\lambda_i$.
Let us define $1/\tilde{\lambda}_i$  the eigenvalues of $\tilde{\Sigma}^{-1}$. 
The following inequalities 
hold \citep[see e.g.][theorem 17, pp. 64-66]{Gantmacher1960}
\begin{equation*}
\frac{1}{\tilde{\lambda}_1}\leq\frac{1}{\lambda_1}\leq\frac{1}{\tilde{\lambda}_2}\leq\frac{1}{\lambda_2}\leq\dots\leq\frac{1}{\tilde{\lambda}_n}\leq\frac{1}{\lambda_n} \; .
\end{equation*}
Because the matrix $\Sigma^{-1}$ is positive-definite, ${1}/{\lambda_i}$ are all positive, thus $\tilde{\Sigma}^{-1}$ has $n-1$ positive eigenvalues. 
Also having $\tilde{\Sigma}^{-1}$ a positive determinant (cf. equation \eqref{eqn:matrixdetlemma}), we conclude that it is positive-definite and condition \eqref{eqn:conditiontheta} is necessary and sufficient to have the whole problem well defined.
In this case, 
after a completion of the square,
we get
\begin{equation*}
\mathbb{E}\left[\exp(\theta V_\textbf{a}({\bf{X}}))\right]=\dfrac{1}{\sqrt{\det(\Sigma\tilde{\Sigma}^{-1})}}\exp\left(-\theta\,\textbf{a}^T\bm{\mu}+\frac{1}{2}\theta^2\,\textbf{a}^T\tilde{\Sigma}\,\textbf{a}\right) \; .
\end{equation*}

\bigskip

Second, 
we consider $\tilde{f}(\textbf{X})$,  the density of $\textbf{X}$ in the alternative model.
For any $\textbf{a}\in\mathcal{A}$, it is
\begin{equation*}
\tilde{f}(\textbf{X})=m^\star_{\theta,\textbf{a}}(\textbf{X})f(\textbf{X}) \; ,
\end{equation*}
which is well-defined if and only if 
$m^\star_{\theta,\textbf{a}}(\textbf{X})$ is well defined. 
In this case, $\tilde{f}(\textbf{X}) $  is a Gaussian density 
with mean and variance \eqref{eqn:mutilde_Sigmatilde} $ \qquad\hspace*{\fill}  \clubsuit \;$

\bigskip

We notice that in the best-case approach, with $\theta<0$, it is not necessary to impose any additional condition for $\theta$; i.e., the alternative measure is well defined $\forall \theta \in \Re^-$:
this is the only difference that should be considered when dealing with the best-case approach.

Condition \eqref{eqn:conditiontheta} determines the domain with all possible values of $\theta$ that allow a well-posed problem, not limited to only small values of $\theta$ and to asymptotic results, as in \citet[][p.31]{Glasserman2014}.
In the rest of this paper, we consider $\theta$ and $\textbf{a}$ in the domain $\mathcal{D}$ defined as
\begin{equation}
\label{eqn:D}
\mathcal{D}:=\left\{ (\theta,\textbf{a}) \text{ s.t. } \theta \, \textbf{a}^T \Sigma \, \textbf{a} < \frac{1}{\gamma} \right\} \, . 
\end{equation} 

\bigskip

We now consider the two external optimisation problems in \eqref{eqn:probLagrange_infinfsup}. 
First, let us define the Lagrangian function 
computed in the optimal change of measure
\begin{equation}
\label{eqn:L_thetaa}
\mathcal{L}(\theta,\textbf{a}):=\mathcal{L}\left(\theta,\textbf{a}; m^\star_{\theta,\textbf{a}}\right)=\frac{1}{\theta} \ln{\mathbb{E}\left[\exp(\theta V_{\bf{a}}({\bf{X}}))\right]} + \frac{\eta}{\theta} \, .
\end{equation}
obtained substituting the optimal change of measure \eqref{eqn:mstarthetaa} in \eqref{eqn:probLagrange_infinfsup}, with $V_{\textbf{a}}(\textbf{X})$ defined in \eqref{eqn:VaX}. Thus, the optimisation problem to be solved becomes
\begin{equation}
\label{eqn:probLagrange_infinf}
\inf_{\textbf{a}}\inf_{\theta>0} \mathcal{L}(\theta,\textbf{a}) \, .
\end{equation} 
The standard technique to solve this problem is to exchange the order of the other two minimisation problems in \eqref{eqn:probLagrange_infinf}. Before entering into details,
we state some properties for the Lagrangian function in the following lemma. 

\begin{lemma}
\label{lem:Lconvex}
Let $(\theta, \textbf{a})\in \mathcal{D}$. 
In the alternative worst-case approach,
$\mathcal{L}(\theta,\textbf{a})$ is convex in $\textbf{a}$ and it has a unique minimum in $\theta$, which is an interior point of the set \eqref{eqn:conditiontheta}; moreover, in the alternative model, the relative entropy (\ref{eqn:relentr_mstar}) becomes
\begin{equation}
\label{eqn:R_thetaa}
R(\theta,\textbf{a}) = \frac{\theta}{2} \, S \; \Gamma( S; \theta, \gamma) + \frac{1}{2}\ln{\left(1-\theta \gamma  \, S \right)} \; ,
\end{equation}
where
\begin{equation}
\label{eqn:S}
S:=\textbf{a}^T\Sigma \, \textbf{a}
\end{equation}
and
\begin{equation}
\label{eq:Gamma}
\Gamma(S; \theta, \gamma) := \dfrac{\gamma \, (1-\theta\gamma \, S)+\theta}{(1-\theta\gamma \, S)^2} \; .
\end{equation}
\end{lemma}

{\bf Proof.} 
First, we prove that the Lagrangian function in the alternative worst-case approach (i.e. with $\theta > 0$)
is convex in $\textbf{a}$, s.t. $(\theta, \textbf{a})\in \mathcal{D}$. We  can apply \citet{ShermanMorrison} formula to get
\begin{equation}
\label{eqn:ShermanMorrison}
\tilde{\Sigma}=\Sigma+\dfrac{\theta\gamma\,\Sigma\,\textbf{a}\textbf{a}^T\Sigma}{1-\theta\gamma\,\textbf{a}^T\Sigma\,\textbf{a}} \; ;
\end{equation}
then, we obtain
\begin{equation}
\label{eqn:EexpthetaVa}
\mathbb{E}\left[\exp(\theta V_\textbf{a}({\bf{X}}))\right] = \dfrac{1}{\sqrt{1-\theta\gamma\,\textbf{a}^T\Sigma\,\textbf{a}}} \exp\left( -\theta\,\textbf{a}^T\bm{\mu} + \frac{1}{2}\theta^2\,\dfrac{\textbf{a}^T\Sigma\,\textbf{a}}{1-\theta\gamma\,\textbf{a}^T\Sigma\,\textbf{a}} \right) \, .
\end{equation}
The Lagrangian function in \eqref{eqn:L_thetaa} becomes
\begin{equation}
\label{eqn:L_thetaa_extended}
\mathcal{L}(\theta,\textbf{a}) = -\frac{1}{2\theta}\ln\left(1-\theta\gamma\,\textbf{a}^T\Sigma \, \textbf{a}\right) -\textbf{a}^T\bm{\mu} + \frac{1}{2}\dfrac{\theta \, \textbf{a}^T\Sigma \, \textbf{a}}{1-\theta\gamma\,\textbf{a}^T\Sigma \, \textbf{a}} +\frac{\eta}{\theta} \, .
\end{equation}
Recalling that the sum of convex functions is itself a convex function \citep[see e.g.][Section 3.2.1, p.79]{BoydConvexOptim}, we focus on the non-linear terms in $\textbf{a}$ in \eqref{eqn:L_thetaa_extended}.
We define
\begin{equation*}
L(\theta,S) := -\frac{1}{2\theta}\ln\left(1-\theta\gamma\,S\right) + \frac{1}{2}\dfrac{\theta \, S}{1-\theta\gamma\,S} \, ,
\end{equation*}
where $S$ is defined in \eqref{eqn:S}.

It is easy to show that $L(\theta,S)$ is convex and increasing in $S$ and that $S$ is convex in $\textbf{a}$ for $\theta >0$. 
Thus, using the composition rule for convexity \citep[see e.g.][Section 3.2.4, p.84]{BoydConvexOptim}, we can conclude that 
$L(\theta,\textbf{a}^T\Sigma \, \textbf{a}) $  is convex in $\textbf{a}$ and then 
the Lagrangian function \eqref{eqn:L_thetaa_extended} is convex in $\textbf{a}$.

\bigskip

Second, we prove the expression \eqref{eqn:R_thetaa} for $R(\theta,\textbf{a})$.
From \eqref{eqn:relentr_mstar}, we have 
\begin{equation}
\label{eqn:relentr_general}
R(\theta,\textbf{a})=\mathbb{E}\left[m^\star_{\theta,\textbf{a}}(\textbf{X})\ln{m^\star_{\theta,\textbf{a}}(\textbf{X})}\right] \; .
\end{equation}
After some simplifications, we get
\begin{equation*}
R(\theta,\textbf{a}) = \frac{\gamma}{2} \, \theta\,\mathbb{E}[(\textbf{a}^T\tilde{\textbf{X}})^2] - \theta \, \mathbb{E}[ \textbf{a}^T \tilde{\textbf{X}} ] + \ln{\sqrt{\det(\Sigma\tilde{\Sigma}^{-1})}} - \frac{1}{2}\,\theta^2\,\textbf{a}^T\tilde{\Sigma}\,\textbf{a}
\end{equation*}
where $\tilde{\textbf{X}}$ is a Gaussian r.v. with mean $- \theta \, \tilde{\Sigma} \, \textbf{a}$ and variance \eqref{eqn:mutilde_Sigmatilde}. 
Finally, using  equation \eqref{eqn:ShermanMorrison} and substituting \eqref{eqn:S} and \eqref{eq:Gamma}, the relative entropy \eqref{eqn:R_thetaa} follows.

\bigskip

Finally, we prove that the Lagrangian function for any given $\textbf{a}$ has a unique minimum in $\theta$, 
called $\tilde{\theta}$, and, in particular, that it is a monotone decreasing function in $(-\infty,\tilde{\theta})$ and a monotone increasing function in $(\tilde{\theta},\theta_{\text{max}}(\textbf{a}))$.

Because $m^\star_{\theta,\textbf{a}}$ is optimal for the Lagrangian function in \eqref{eqn:probLagrange_infinfsup} and 
the alternative model $\tilde{f} (\bf{X})$ 
is a sufficiently regular function (a multinomial Gaussian p.d.f.), 
we can exchange the derivative with the expected value \citep[cf. e.g., ][Th.4, p.429]{Protter} and 
we get
\begin{equation*}
\frac{\partial \mathcal{L} (\theta,\textbf{a}) }{\partial \theta}  = \frac{\partial}{\partial \theta} \left\{ \mathbb{E}\left[m^\star_{\theta,\textbf{a}}({\bf{X}})V_\textbf{a}({\bf{X}})-\frac{1}{\theta} \bigg( m^\star_{\theta,\textbf{a}}({\bf{X}})\ln{m^\star_{\theta,\textbf{a}}({\bf{X}})}-\eta \bigg) \right] \right\} = \frac{1}{\theta^2} \left( R(\theta,\textbf{a})-\eta \right) \, .
\end{equation*}

Observing that the relative entropy $R(\theta,\textbf{a})$ is null in $\theta=0 \, \forall \textbf{a} $, 
it is monotone increasing for $\theta >0$ (cf. Lemma \ref{lem:relentr_monotone}) and 
it tends to infinity if $\theta\rightarrow \theta_{\text{max}}(\textbf{a})$ (straightforward using \eqref{eqn:R_thetaa} and \eqref{eq:Gamma}), 
the thesis follows $ \qquad\hspace*{\fill}  \clubsuit \; $

\bigskip

\citet{Glasserman2014} prove that, under certain conditions, it is possible to exchange the two infima in \eqref{eqn:probLagrange_infinf} 
and subsequently they solve the optimisation problem 
 in the variables $\textbf{a}$ and $\theta$ numerically.
We use the same inversion as \citet{Glasserman2014}, but we solve analytically the problem: as already stated in the introduction, this is our main theoretical contribution.

To find the optimal portfolio in the alternative measure and the corresponding optimal change of measure, we recall Proposition 2.1 of 
\citet{Glasserman2014} for the portfolio selection.

\begin{lemma}
\label{lem:astar_mstartheta}
Let $(\theta, \textbf{a})\in \mathcal{D}$.
The optimisation problem \eqref{eqn:probLagrange_infinf} is equivalent to 
\begin{equation}
\label{eqn:probLagrange2}
\inf_{\theta>0} \inf_{{\bf{a}} \in \mathcal{A}} \mathcal{L}(\theta,\textbf{a})
\end{equation}
where $\mathcal{L}(\theta,\textbf{a})$ is defined in \eqref{eqn:L_thetaa};
the optimal portfolio ${\bf{a}}^\star(\theta)$ in the alternative measure is
\begin{equation}
\label{eqn:astar}
{\bf{a}}^\star(\theta)=\arg\inf_{{\bf{a}} \in \mathcal{A}} \frac{1}{\theta} \ln{\mathbb{E}\left[\exp(\theta V_{\bf{a}}({\bf{X}}))\right]}
\end{equation}
and the worst-case change of measure is
\begin{equation}
\label{eqn:mstartheta}
m^\star_{\theta}({\bf{X}})=\dfrac{\exp\left(\theta V_{{\bf{a}}^\star(\theta)}({\bf{X}})\right)}{\mathbb{E}\left[\exp\left(\theta V_{{\bf{a}}^\star(\theta)}({\bf{X}})\right)\right]} \; .
\end{equation}
\end{lemma}

{\bf Proof.}
See  Proposition 2.1 in \citet[][]{Glasserman2014}, 
applied to the mean-variance case    
$ \qquad\hspace*{\fill}  \clubsuit \; $

\bigskip


This lemma is an important result known from literature about the worst-case model risk.
As in \citet{Glasserman2014}, we follow the same approach in this paper. 
We first find the optimal worst-case portfolio ${\bf a}$ given $\theta>0$ and then select the optimal value of $\theta$ within the KL-ball $P_\eta$. 
It is straightforward to prove that in the best-case approach the same equations of Lemma \ref{lem:astar_mstartheta} hold with $\theta < 0$.

In the rest of this paper, we find an analytical solution for \eqref{eqn:probLagrange2} in two cases: 
the general mean-variance framework in Section ${\bf 3}$ and the special case considered by \citet{Glasserman2014} with a constraint on the mean in the alternative model,
analysed in Section ${\bf 4}$.



\section{Analytical solution for the worst-case portfolio selection}
In this section, we solve analytically the optimisation problem \eqref{eqn:astar} and find the optimal portfolio in the alternative model for a given value of $\theta>0$; 
i.e., the robust portfolio in the mean-variance framework. 
Then, we prove that the optimum $\theta$ in  \eqref{eqn:probLagrange2} stays on the surface of the ball $P_\eta$. 


\begin{theorem}
\label{thm:astartheta_meanvariance}
Let $(\theta, \textbf{a})\in \mathcal{D}$. In the alternative worst-case approach,
the optimal portfolio ${\bf{a}}^\star(\theta)$ is
\be
\label{eqn:system_astartheta}
\displaystyle {\bf{a}}^\star(\theta)=
\frac{A}{\Gamma(S^\star(\theta); \theta, \gamma)} \; \frac{\Sigma^{-1}\bm{\mu}}{A} + 
\left(1- \frac{A}{\Gamma(S^\star(\theta); \theta, \gamma)}  \right)\frac{\Sigma^{-1}\textbf{1}}{C} 
\en
and the quantity $S^\star(\theta)$ 
is the unique solution of
\be
 \displaystyle S=\frac{1}{C} \; \left( \frac{D}{\Gamma(S; \theta, \gamma)^2}+ 1 \right)
\label{eqn:system_astartheta2}
\en
where $\Gamma(S; \theta,\gamma)$ is defined in \eqref{eq:Gamma}.
The constants $A, C, D$ have been defined in \eqref{eqn:ABCD}. 
\end{theorem}

{\bf Proof.}
In the alternative worst-case approach, it is possible to find the optimal portfolio solving \eqref{eqn:astar}.
Using \eqref{eqn:EexpthetaVa} and introducing a Lagrange multiplier for the constraint ${\bf{a}} \in \mathcal{A}$ (i.e. ${\bf{a}}^T {\bf{1}} = 1$), 
\eqref{eqn:astar} is equivalent to
\begin{equation*}
\arg\inf_{\textbf{a}} -\frac{1}{2\theta} \ln{ \left(1-\theta\gamma\,\textbf{a}^T\Sigma\,\textbf{a}\right) } - \textbf{a}^T\bm{\mu} + \frac{1}{2}\theta\,\dfrac{\textbf{a}^T\Sigma\,\textbf{a}}{1-\theta\gamma\,\textbf{a}^T\Sigma\,\textbf{a}} + \alpha \left(1-\textbf{a}^T\textbf{1} \right) .
\end{equation*}
Following the same method as in the nominal model,
we get the equation \eqref{eqn:system_astartheta} for $\textbf{a}$,
where $S$ is defined in \eqref{eqn:S}.
By substituting \eqref{eqn:system_astartheta} in \eqref{eqn:S}, we get the equation \eqref{eqn:system_astartheta2} for the optimal $S$.

\bigskip

Let us study the solutions of equation  \eqref{eqn:system_astartheta2} given $\Gamma(S; \theta, \gamma)$ in   \eqref{eq:Gamma}. The domain of $S$ is s.t. $S$ must be included in the interval
\begin{equation}
\label{eq:S domain}
\frac{1}{C}\leq S<\frac{1}{\theta\gamma} \; ,
\end{equation}
because i) $S={1}/{C}$ corresponds to the minimum-variance case
even in the alternative model 
and ii) $(\theta, \textbf{a})\in \mathcal{D}$, i.e. $\theta S < 1/\gamma$. 

\begin{figure}
\centering
{\includegraphics[width=.55\textwidth]{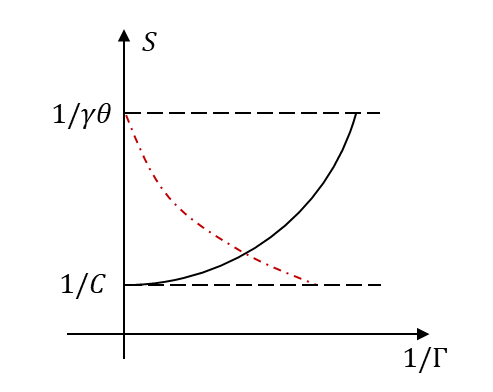}} 
\caption{\small 
Plot of $S$ as a function of ${1}/{\Gamma}$ in equation \eqref{eqn:system_astartheta2} (continuous black line) and 
${1}/{\Gamma}$  as a function of $S$ in equation \eqref{eq:Gamma} for $S\in [{1}/{C}, {1}/{\theta\gamma} )$ (dot-dashed red line).
The system of two equations admits a unique solution.}
\label{fig:Soluzione}
\end{figure}

First, let us consider $S$ as a function of ${1}/{\Gamma}$ in equation \eqref{eqn:system_astartheta2}
in the domain \eqref{eq:S domain}, which is one branch of a parabola (and then monotone increasing) with a minimum in $(0, 1/C)$. 
Then, consider ${1}/{\Gamma}$  as a function of $S$ in equation \eqref{eq:Gamma} in the same domain for $S$,
which is equal to a positive value in $S= 1/C$ and it is derivable in the domain of $S$ with 
 a derivative always strictly negative and it tend to zero in the limit $S \to 1/ (\theta\gamma)$.
Hence, Equation \eqref{eqn:system_astartheta2} has a unique solution, as shown in Figure \ref{fig:Soluzione} $ \qquad\hspace*{\fill}  \clubsuit \;$


\bigskip

Depending on $\theta$, the worst-case mean-variance optimal portfolio is
the (unique) analytical solution in equation \eqref{eqn:system_astartheta2}. 
Let us observe that, 
in the mean-variance framework, the optimal worst-case portfolio is similar to the optimal nominal one \eqref{eqn:astarclassic}. 
Also in the worst-case approach  a two mutual fund theorem holds: the optimal portfolio is the linear combination --with a different weight-- of
the same two portfolios 
$\textbf{a}_1^\star$ and $\textbf{a}_0^\star$ of the nominal problem.
The solution in the alternative model has exactly the same form of the nominal one with an increased risk aversion ``parameter" $\Gamma$ in the worst-case approach ($\Gamma > \gamma$ for $\theta > 0)$.
It can be shown that a similar solution holds also in the best-case approach  with
a decreased risk aversion ($\Gamma < \gamma$ for $\theta < 0$).

Let us notice that, given the optimal portfolio $\textbf{a}^\star(\theta)$, we are now able to find the corresponding relative entropy, i.e. the KL-divergence between the nominal and the alternative model, depending only on parameter $\theta$. Substituting the unique solution of \eqref{eqn:system_astartheta2} in \eqref{eqn:R_thetaa}, we get
\begin{equation}
\label{eqn:relentr}
R(\theta) := R(\theta,  {\bf{a}}^\star(\theta)) = \frac{\theta}{2} \, S^\star(\theta) \; \Gamma( S^\star(\theta); \theta, \gamma) + \frac{1}{2}\ln{\left(1-\theta \gamma  \, S^\star(\theta) \right)} \; .
\end{equation}

We conclude this section proving that the optimal parameter $\theta$ in the optimisation problem \eqref{eqn:probLagrange2} stays on the surface of the ball $P_\eta$. 
This result is crucial because solves completely \eqref{eqn:probLagrange2}, i.e. the  mean-variance portfolio selection in the worst-case approach.

\begin{proposition}
\label{prop:thetastarsurface}
Let $(\theta, \textbf{a}^\star (\theta))\in \mathcal{D}$.
The optimal value $\theta^\star$ for $\theta$ in optimisation problem \eqref{eqn:probLagrange2} is on the surface of the ball $P_\eta$, i.e. $R(\theta^\star) = \eta$ 
where $R(\theta)$ is defined in \eqref{eqn:relentr}.
\end{proposition}

{\bf Proof.}
We first show that the optimum $\theta$ that solves the original optimisation problem \eqref{eqn:probLagrange_infinfsup} is on the surface of the ball $P_\eta$ for any given portfolio $\textbf{a}$; 
then, we consider
problem \eqref{eqn:probLagrange2} obtained inverting the two infima.

\bigskip

First, we consider the internal maximisation problem in \eqref{eqn:probopt1infsup} for a given $\textbf{a}$, that is the primal problem.
Following the 
notation in \citet[][pp.216-225]{BoydConvexOptim}
we call the primal problem $p$, 
we  indicate with $d$ the corresponding Lagrangian dual problem in \eqref{eqn:probLagrange_infinfsup},
while $p^\star$ and $d^\star$ denote the optimal values respectively of the primal problem and of the dual one. 

Because the objective function is convex in $m$ and the set $\left\{m: \mathbb{E}[m(\textbf{X}) \ln m(\textbf{X}) ]<\eta \right\}$ is non-empty, 
Slater's theorem ensures that strong duality holds, i.e. $p^\star=d^\star$ \citep[see, e.g.][Section 5.2.3, p.226]{BoydConvexOptim}. 
In other words, for a given $\textbf{a}$,
given $m^{\star p}_{\textbf{a}}(\textbf{X})$ a primal optimum and $\{ \tilde{\theta}, m^{\star}_{\tilde{\theta}, {\textbf{a}}}(\textbf{X}) \}$ a dual optimum
where $\tilde{\theta}$ is a function of $\textbf{a}$, we have
\begin{align*}
p^\star &= d^\star = \inf_{\theta>0} \sup_{m} \mathbb{E}\left[ m(\textbf{X})V_\textbf{a}(\textbf{X}) - \frac{1}{\theta} \left( m(\textbf{X}) \ln m(\textbf{X}) -\eta \right)\right] = \\
&= \sup_{m} \inf_{\theta>0} \mathbb{E}\left[ m(\textbf{X})V_\textbf{a}(\textbf{X}) - \frac{1}{\theta} \left( m(\textbf{X}) \ln m(\textbf{X}) -\eta \right)\right] = \\
&= \sup_{m} \mathbb{E}\left[ m(\textbf{X})V_\textbf{a}(\textbf{X}) - \frac{1}{\tilde{\theta}} \left( m(\textbf{X}) \ln m(\textbf{X}) -\eta \right)\right] \geq \\
& \geq \mathbb{E}\left[ m^{\star p}_{\textbf{a}}(\textbf{X})V_\textbf{a}(\textbf{X}) - \frac{1}{\tilde{\theta}} \left( m^{\star p}_{\textbf{a}}(\textbf{X}) \ln m^{\star p}_{\textbf{a}}(\textbf{X}) -\eta \right)\right] \geq \\
&\geq \mathbb{E} \left[ m^{\star p}_{\textbf{a}}(\textbf{X})V_{\textbf{a}}(\textbf{X}) \right] = p^\star \, .
\end{align*}
The second line comes from \citet[][eq.(5.45), p.238]{BoydConvexOptim}. 
The third line considers the optimum $\tilde{\theta}$ of the dual problem. 
The fourth line follows because the supremum of the Lagrangian over $m$ is greater or equal to its value at any other $m(\textbf{X})$ and then also choosing
$m(\textbf{X})=m^{\star p}_{\textbf{a}}(\textbf{X})$. 
The last inequality is due to the fact that the second term in fourth line is non-negative.

Thus, since all inequalities hold with equality, we can draw two conclusions. First, $m^{\star p}_{\textbf{a}}(\textbf{X})$ maximises the Lagrangian. This result, added to the concavity of the Lagrangian in $m$, implies that $m^{\star p}_{\textbf{a}}(\textbf{X}) = m^{\star}_{\tilde{\theta}, {\textbf{a}}}(\textbf{X})$.

Second, the following equality holds
\begin{equation}
\label{eqn:thetastar}
\frac{1}{\tilde{\theta}} \, \mathbb{E} \left[ \left( m^{\star}_{\tilde{\theta}, {\textbf{a}}}(\textbf{X}) \ln m^{\star}_{\tilde{\theta}, {\textbf{a}}}(\textbf{X}) -\eta \right)\right]=0 \, .
\end{equation}
Because $\theta$ must remain finite, due to condition \eqref{eqn:conditiontheta}, we get that the expectation in \eqref{eqn:thetastar} must be null; 
i.e., $\tilde{\theta}$ stays on the surface of the ball $P_\eta$, or equivalently $R(\tilde{\theta}, \textbf{a}) = \eta $ $\forall \textbf{a}$ s.t. $(\tilde{\theta}, \textbf{a})\in \mathcal{D}$.

Because i) the relative entropy $R(\theta, \textbf{a}) $ is a monotone increasing function in $\theta$ (cf. Lemma \ref{lem:relentr_monotone}), 
ii) it is null when $\theta=0$ and iii) it tends to infinity in the limit $\theta \rightarrow \theta_{\text{max}}(\textbf{a})$ (cf. also the end in the proof of Lemma \ref{lem:Lconvex}), 
then there exists a unique solution of \eqref{eqn:thetastar} for $\tilde{\theta} \in \left[0,\theta_{\text{max}}(\textbf{a}) \right)$ and then 
also a unique solution for the dual optimum  $\{ \tilde{\theta}, m^{\star}_{\tilde{\theta}, {\textbf{a}}}(\textbf{X}) \}$.

\bigskip


In Lemma \ref{lem:astar_mstartheta} we have proven that it is possible to change the order of the two infima in \eqref{eqn:probLagrange_infinfsup}, obtaining the same solution. 
This is the inverted dual problem \eqref{eqn:probLagrange2}.
Thus, the optimum value $\theta^\star$ 
of the inverted dual problem is as well on the surface of the ball, that ends the proof
$ \qquad\hspace*{\fill}  \clubsuit \;$

\bigskip

The result proved in Proposition \ref{prop:thetastarsurface} 
allows us 
i) to avoid the numerical optimisation in the parameter $\theta$ and 
ii) to identify the optimal $\theta$, corresponding to the divergence $\eta$, as the (unique) positive value on the surface of the ball $P_\eta$.
This completely solves the problem \eqref{eqn:probLagrange2}.

\section{Mean-variance with constant mean}
In this section we analyse in detail the mean-variance case considered by \citet{Glasserman2014}, in which they consider a special case with a constraint on the mean vector that must be equal to the nominal mean $\bm{\mu}$ even in the alternative model. The problem in this case becomes
\begin{equation*}
\begin{cases}
	\inf_{\textbf{a}\in\mathcal{A}(\theta)}\sup_{m\in P_\eta}\mathbb{E} \left [m({\textbf{X}}) \left( \dfrac{\gamma}{2}\textbf{a}^T({\textbf{X}}-\bm{\mu})({\textbf{X}}-\bm{\mu})^T\textbf{a}-\textbf{a}^T\textbf{X} \right) \right] \\ 
	\text{s.t. } \mathbb{E}\left[ m({\textbf{X}}) \cdot {\textbf{X}} \right]=\bm{\mu}
\end{cases}
\end{equation*}
as in \citet[][p.36]{Glasserman2014}, that is equivalent to
\begin{equation*}
\inf_{\textbf{a}\in\mathcal{A}(\theta)}\sup_{m\in P_\eta}\mathbb{E} 
\left [m({\textbf{X}}) V_{\textbf{a}}^{GX}(\textbf{X}) \right]
\end{equation*}
with the measure of risk $V_{\textbf{a}}^{GX}(\textbf{X})$ defined in \eqref{eqn:VaGX_X}.

\bigskip

In the remaining part of this section, we show that all results proved in previous sections can be replicated in this special case.

First, let us show that the basic properties of Lemmas \ref{lem:ftilde}, \ref{lem:Lconvex} and \ref{lem:astar_mstartheta}
can be adapted to this special case, considering the measure of risk \eqref{eqn:VaGX_X}. This is proven in the next Lemmas  \ref{lem:GX_ftilde} and \ref{lem:GX_Lconvex}.

\begin{lemma}
Let $(\theta, \textbf{a})\in \mathcal{D}$.
The change of measure $m^\star_{\theta,{\bf{a}}}(\textbf{X})$ in \eqref{eqn:mstarthetaa} is well-defined if and only if condition \eqref{eqn:conditiontheta} holds.
Moreover,  for any ${\bf{a}}\in\mathcal{A}$,  in the alternative model $\tilde{f}(\textbf{X})$ corresponding to $m^\star_{\theta,{\bf{a}}}(\textbf{X})$,
$\textbf{X}$ is distributed as a multivariate normal r.v., i.e. $\textbf{X}\sim N(\bm{\mu},\tilde{\Sigma})$, with $\tilde{\Sigma}$ as in \eqref{eqn:mutilde_Sigmatilde}. 
\label{lem:GX_ftilde}
\end{lemma} 

{\bf Proof.}
See proof of Lemma \ref{lem:ftilde}, noting that $\mathbb{E}\left[\exp\left( \theta V_{\textbf{a}}^{GX}(\textbf{X}) \right)\right]$ attains the same value as in mean-variance framework with a linear constant term in the exponential $-\theta \, \textbf{a}^T\bm{\mu}$ instead of the linear stochastic term $-\theta \, \textbf{a}^T\textbf{X}$. Thus, the first part of the proof holds and it is straightforward to show that $\tilde{f}(\textbf{X})$ is the density of a Gaussian r.v. with mean $\bm{\mu}$ (the new constraint that we are imposing) and variance $\tilde{\Sigma}$  $ \qquad\hspace*{\fill}  \clubsuit \; $

\bigskip

Defining $\mathcal{L}^{GX}(\theta,\textbf{a})$ as in \eqref{eqn:L_thetaa}, using $V_\textbf{a}^{GX}(\textbf{X})$ instead of $V_\textbf{a}(\textbf{X})$, 
we can show that results similar to Lemma \ref{lem:Lconvex} and Lemma \ref{lem:astar_mstartheta} hold as well.

\begin{lemma}
Let $(\theta, \textbf{a})\in \mathcal{D}$.
$\mathcal{L}^{GX}(\theta,\textbf{a})$ is convex in $\textbf{a}$ and it has a unique minimum in $\theta$, interior point of the set \eqref{eqn:conditiontheta}. In the alternative worst-case approach, the relative entropy is 
\begin{equation}
\label{eqn:relentr_GX}
R(\theta,\textbf{a})=\frac{\theta}{2} \, S \; \Gamma^{GX}( \theta, \gamma) + 
\frac{1}{2}\ln{\left(1-\theta \; \gamma  \; S \right)} \; ,
\end{equation}
where 
\begin{equation}
\label{eqn:GammaGX}
\Gamma^{GX}(S;\theta,\gamma):=\frac{\gamma}{1-\theta\gamma \, S} \, .
\end{equation}
Moreover, it is possible to exchange the order of the two inferior in the optimisation problem \eqref{eqn:probLagrange_infinfsup}, that becomes equivalent to
\begin{equation*}
\inf_{\theta>0}\inf_{\textbf{a}\in\mathcal{A}} \mathcal{L}^{GX}(\theta, \textbf{a}) \, ;
\end{equation*}
the optimal portfolio in the alternative measure is found solving 
\begin{equation}
\label{eqn:astar_VaGX}
{\bf{a}}^\star(\theta)=\arg\inf_{{\bf{a}} \in \mathcal{A}} \frac{1}{\theta} \ln{\mathbb{E}\left[\exp(\theta V_{\bf{a}}^{GX}({\bf{X}}))\right]} \, .
\end{equation}
 \label{lem:GX_Lconvex}
\end{lemma}

{\bf Proof.}
See proof of Lemma \ref{lem:Lconvex} and Lemma \ref{lem:astar_mstartheta}, noting that
\begin{equation*}
\mathcal{L}^{GX}(\theta,\textbf{a}) = -\frac{1}{2\theta}\ln\left(1-\theta\gamma\,S\right) -\textbf{a}^T\bm{\mu} +\frac{\eta}{\theta}
\end{equation*} 
and for the relative entropy that
\begin{equation*}
R(\theta,\textbf{a})=\mathbb{E}\left[ m^\star_\theta(\textbf{X})\frac{\theta\gamma}{2}\textbf{a}^T(\textbf{X}-\bm{\mu})(\textbf{X}-\bm{\mu})^T\textbf{a} \right] -\ln \mathbb{E}\left[ \exp\left(\frac{\theta\gamma}{2}\textbf{a}^T(\textbf{X}-\bm{\mu})(\textbf{X}-\bm{\mu})^T\textbf{a} \right) \right] \, .
\end{equation*}
Following similar computations as in mean-variance case, all previous results hold $ \qquad\hspace*{\fill}  \clubsuit \; $

\bigskip

Then, similarly to Theorem \ref{thm:astartheta_meanvariance} we can find the
closed form solution of problem \eqref{eqn:astar_VaGX} for a given positive $\theta$ and prove that the optimum stays on the surface of the ball $P_\eta$ also in this special case. 

\begin{theorem}
\label{thm:astar_GX}
Let $(\theta, \textbf{a})\in \mathcal{D}$. 
In the alternative worst-case approach, the optimal portfolio is
\be
\textbf{a}^\star(\theta)=\frac{A}{\Gamma^{GX}(\theta, \gamma)} \; \frac{\Sigma^{-1}\bm{\mu}}{A} + 
\left(1- \frac{A}{\Gamma^{GX}(\theta, \gamma)}  \right)\frac{\Sigma^{-1}\textbf{1}}{C} 
\label{eqn:system_astartheta_GX}
\en
where 
\begin{equation}
\label{eqn:Gammastar_GX}
\Gamma^{GX}(\theta, \gamma)=\dfrac{\gamma C + \sqrt{\gamma^2 C^2+4\theta\gamma \, (C-\theta\gamma)D}}{2 \, (C-\theta\gamma)} \, ,
\end{equation}
with $\Gamma^{GX}(\theta, \gamma) >0$
\end{theorem}

{\bf Proof.}
As in the mean-variance case, it is possible to find the optimal portfolio in the alternative measure solving \eqref{eqn:astar_VaGX}, that becomes
\begin{equation*}
\inf_{{\bf{a}}:{\bf{a}}^T\mathbf{1}=1 } \; \frac{1}{\theta} \ln{\mathbb{E}\left[\exp\left( \frac{\theta\gamma}{2} \, \textbf{a}^T(\textbf{X}-\bm{\mu})(\textbf{X}-\bm{\mu})^T \textbf{a}\right)\right]} -{\bf{a}}^T \bm{\mu} + \frac{\eta}{\theta}
\end{equation*} 
that, introducing a Lagrange multiplier for the constraint $\textbf{a}^T\textbf{1}=1$, is equivalent to
\begin{equation*}
\inf_{{\bf{a}}} -\frac{1}{2\theta}\ln{\left(1-\theta\gamma\,\textbf{a}^T\Sigma\,\textbf{a}\right)}-{\bf{a}}^T \bm{\mu} + \frac{\eta}{\theta} -\lambda({\bf{a}}^T\mathbf{1}-1) \, .
\end{equation*}
Following the same method as in the nominal model, we get equation \eqref{eqn:system_astartheta2} with $\Gamma^{GX}(S;\theta,\gamma)$ defined in \eqref{eqn:GammaGX} 
instead of $\Gamma(S; \theta,\gamma)$ defined in \eqref{eq:Gamma}.

As in the mean-variance case, the domain for $S$ is \eqref{eq:S domain}.
Moreover, in this special case, equation \eqref{eqn:system_astartheta2} and \eqref{eqn:GammaGX} represent a second order degree system, that can be solved in closed form.

We obtain the following solution for a second order degree equation in the variable $\Gamma^{GX}$
\begin{equation*}
\Gamma^{GX}(\theta,\gamma)=\dfrac{\gamma C \pm \sqrt{\gamma^2 C^2+4\theta\gamma \, (C-\theta\gamma)D}}{2 \, (C-\theta\gamma)} \, . 
\end{equation*}
Observing from \eqref{eq:S domain} that $C>\theta\gamma$ and from \eqref{eqn:GammaGX} that $\Gamma^{GX}(S;\theta,\gamma)>0$,
we can deduce that the solution for $\Gamma^{GX}(\theta,\gamma)$ with the negative sign is non-acceptable in the worst-case approach, 
 thus \eqref{eqn:Gammastar_GX} is the  unique solution for $\Gamma^{GX}(\theta,\gamma)$,
that concludes the proof  
$ \qquad\hspace*{\fill}  \clubsuit \; $

\bigskip

As in the mean-variance case, let us notice that, given the optimal portfolio \eqref{eqn:system_astartheta_GX}, we are able to obtain from \eqref{eqn:relentr_GX}
 the corresponding relative entropy that depends only on parameter $\theta$; i.e.,
\begin{equation}
\label{eqn:relentr_theta_GX}
R(\theta) = \frac{1}{2} \left( \frac{\Gamma^{GX}(\theta, \gamma)}{\gamma} -1 - \ln \frac{\Gamma^{GX}(\theta, \gamma)}{\gamma} \right) 
\; .
\end{equation}

Also Proposition \ref{prop:thetastarsurface} can be replicated in this special case.

\begin{proposition}
\label{prop:thetastarsurface_GX}
Let $(\theta, \textbf{a}^\star(\theta))\in \mathcal{D}$.
The optimal value $\theta^\star$ for $\theta$ in the worst-case portfolio selection 
 is on the ball, i.e. $R(\theta^\star) = \eta$, with $R(\theta)$ in \eqref{eqn:relentr_theta_GX}.
\end{proposition}

{\bf Proof.}
The proof follows the same steps of Proposition \ref{prop:thetastarsurface} with the Lagrangian and entropy of this special case $ \qquad\hspace*{\fill}  \clubsuit \; $

\bigskip

We conclude the section noting that the value of the risk measure in this framework simply becomes 
\begin{equation}
\label{eqn:exp_utility_minimumvar}
\mathbb{E}[m^\star_\theta(\textbf{X})\,V^{GX}_{\textbf{a}^\star(\theta)}(\textbf{X})]=
\frac{\gamma}{2} \; {\textbf{a}^\star (\theta)}^T\tilde{\Sigma}\,\textbf{a}^\star (\theta) - {\textbf{a}^\star (\theta)}^T\bm{\mu} =
\frac{1}{2 \, C} \left( \Gamma^{GX}(\theta, \gamma) - \frac{D}{\Gamma^{GX}(\theta, \gamma)} \right) - \frac{A}{C}
\end{equation}
where in the second equality we have used \eqref{eqn:ShermanMorrison} and the optimal portfolio $\textbf{a}^\star (\theta)$ in \eqref{eqn:system_astartheta_GX}.

In practice, an efficient way to obtain a graphical representation of the result is
first to identify the alternative model through the parameter $\theta$, 
(e.g. in a range $\left[0,\hat{\theta}\right]$), 
and then to get  the relative entropy  \eqref{eqn:relentr_theta_GX} for that value of $\theta$.  
The numerical examples in next section are carried out in this way, obtaining at the same time the  relative entropy \eqref{eqn:relentr_theta_GX} and
the associated value of the risk measure
\eqref{eqn:exp_utility_minimumvar}.

\section{Equality of optimal portfolio in alternative and nominal models and numerical examples}

In this section, we analyse the cases where the optimal portfolio in the nominal model and in the alternative one are equal.
We prove that this equality holds only in two relevant cases:
the minimum-variance problem in which the uncertainty is limited only to the covariance matrix and 
a symmetric case where all assets have the same mean, i.e. $\bm{\mu}=\mu \; \textbf{1}$.

\bigskip

The minimum-variance case is an interesting subcase of the mean-variance portfolio selection
--as of the mean-variance with fixed mean portfolio selection considered in \citet{Glasserman2014}--
and it is a purely risk-based approach to portfolio construction.
This corresponds to selecting a very large risk adversion parameter $\gamma$ in \eqref{eqn:VaX} or equivalently in \eqref{eqn:VaGX_X}.
In this case, the measure of risk is
\[
V_\textbf{a}(\textbf{X}):=\frac{1}{2} \, \textbf{a}^T(\textbf{X}-\bm{\mu})(\textbf{X}-\bm{\mu})^T \textbf{a} \; .
\]

In the minimum-variance case, we can prove the interesting analytical result that the optimal portfolio in the worst-case approach is exactly the same as the optimal portfolio in the nominal model. Moreover, we can adapt all results obtained in previous Sections to this case.

\bigskip

First, 
we can adapt Lemma \ref{lem:GX_ftilde} 
and Proposition \ref{prop:thetastarsurface_GX}
to the minimum-variance case, obtaining that: 
i) in the alternative model, $\textbf{X}$ is distributed as a multivariate normal r.v. 
with the same mean $\bm{\mu}$ as the nominal model and variance $\tilde{\Sigma}$ as in \eqref{eqn:mutilde_Sigmatilde}, if and only if the same condition \eqref{eqn:conditiontheta}, with $\gamma =1$, holds;
ii) the relative entropy  $R(\theta)$ associated to the optimal worst-case and nominal portfolio has the same expression \eqref{eqn:relentr_theta_GX} as in previous section, with $\gamma=1$, and the optimal $\theta$ is on the surface of the ball $P_\eta$.

Then, we can prove that the optimal portfolio in the alternative model (i.e. the robust portfolio) is the same of the optimal one in the nominal model, as shown in the next proposition.

\begin{proposition}
\label{prop:astartheta_minimumvariance}
Let $(\theta, \textbf{a})\in \mathcal{D}$.
In the minimum-variance case,
the optimisation problem \eqref{eqn:astar} is equivalent to 
\begin{equation}
\label{eqn:astartheta_minvar}
{\bf{a}}^\star(\theta)=\arg\inf_{{\bf{a}}\in\mathcal{A}}{\bf{a}}^T\Sigma\, {\bf{a}}\; ,
\end{equation}
thus, the optimal portfolio in the alternative model $\tilde{f}(\textbf{X})$ is the same as the one of the nominal model $f(\textbf{X})$, i.e. $\textbf{a}^\star(\theta)=\textbf{a}_0^\star$.
\end{proposition}

{\bf Proof.}
The optimal portfolio is found solving optimisation problem \eqref{eqn:astar}. 
After some computations and using \eqref{eqn:matrixdetlemma}, we get
\begin{equation*}
\mathbb{E}\left[\exp(\theta V_\textbf{a}({\bf{X}}))\right]=\dfrac{1}{\sqrt{\det(\Sigma\tilde{\Sigma}^{-1})}}=\dfrac{1}{\sqrt{1-\theta \,\textbf{a}^T\Sigma\,\textbf{a}}} \; .
\end{equation*}
Hence, the worst-case problem \eqref{eqn:astar} is equivalent to the classical problem.
Then, the solution is the same of the nominal model and it is unique  $ \qquad\hspace*{\fill}  \clubsuit \;$

\bigskip


Let us stress that two main properties 
hold in the  minimum-variance framework: 
not only the exponential change of measure stays in the family of multivariate normal distributions, 
but also the robust portfolio is equal to the optimal one in the nominal model.
These two properties lead to the consequence that in the minimal-variance framework
the worst-case approach corresponds to
a change in the parameters of the Gaussian distribution:
in this situation model risk can be explained simply as an {\it estimation} risk.

\bigskip

In the remaining part of this section, we get back to the mean-variance framework and we focus our attention on a symmetric case
\begin{equation*}
\bm{\mu}=\mu \; \textbf{1} =\begin{bmatrix}
\mu \\ \mu \\ \vdots \\ \mu
\end{bmatrix} \; .
\end{equation*}
We first prove a general result of a necessary and sufficient condition for the equality of the optimal portfolio in the nominal and in the alternative model:
in this case model risk is equivalent to an {\it estimation} risk.
Then, we show some numerical examples.


\bigskip

In the general mean-variance framework (and also in the special case), it is possible to prove the following proposition, 
that  guarantees the equality of the optimal portfolios in the nominal and in the alternative model.
\begin{proposition}
\label{prop:necsufcondition}
Let $(\theta, \textbf{a})\in \mathcal{D}$ and  $\gamma$ a finite risk aversion.
The optimal mean-variance portfolio in the nominal and 
alternative models are equal 
if and only if $\bm{\mu}=\mu \, \textbf{1}$ with $\mu \in \mathbb{R}$.
\end{proposition}

{\bf Proof.}
From equations \eqref{eqn:astarclassic}  and \eqref{eqn:system_astartheta}, we have
\begin{equation*}
\textbf{a}^\star(\theta)=\textbf{a}^\star_{\text{nom}} \, \Leftrightarrow \, 
\left(\Sigma^{-1}\bm{\mu} - \frac{A}{C} \, \Sigma^{-1}\textbf{1}\right) \, 
\left(\frac{1}{\gamma}-\frac{1}{\Gamma(S,\theta,\gamma)}\right)
=\textbf{0} \, ,
\end{equation*}
 where $A$ and $C$ have been defined in (\ref{eqn:ABCD}) and $\textbf{0}$ is the null vector in $\mathbb{R}^n$.
The equation has a solution only if either
$\text{i) }    C\Sigma^{-1}\bm{\mu} = A\Sigma^{-1}\textbf{1} \; \text{ or }
\text{ ii) } $
$ \Gamma_2(S,\theta,\gamma)^{-1} =\gamma^{-1} \; \Leftrightarrow \; \theta=0$.
While, for finite $\gamma >0$, the condition ii)  cannot be fulfilled in the alternative model, condition i) proves the proposition:
it corresponds to have $\bm{\mu}$ in the same direction of $\textbf{1}$ with $\mu = A/C$
$ \qquad\hspace*{\fill}  \clubsuit \; $

%

\bigskip

It can be interesting to underline that Proposition \ref{prop:necsufcondition} holds even in the mean-variance case with a constraint on 
the mean vector considered by \citet{Glasserman2014}.

\bigskip

In order to show some numerical examples, 
let us consider, as in \citet{Glasserman2014}, the case of mean-variance with fixed mean and
a fully symmetric variance, i.e.
\begin{equation*}
\Sigma=\sigma^2\begin{bmatrix}
1 & \rho & \dots & \rho \\
\rho & 1 & \dots & \rho \\
\vdots & \vdots & \ddots & \vdots \\
\rho & \rho & \dots & 1
\end{bmatrix} \; ,
\end{equation*}
with $\rho > - 1/(n-1)$.
This case presents the advantage of a complete detailed analytical solution for a generic $n$ 
and it allows to understand in an interesting example what can happen in the numerical determination of the optimal solution in the alternative model. 
The eigenvalues of the variance-covariance matrix $\Sigma$ are (see, e.g. Lemma \ref{lem:eigenSigma} in Appendix A):
\begin{equation}
\label{eqn:eigenvaluesSigma}
\left\{
\begin{array}{ll}
\displaystyle \lambda_1=\sigma^2(1+(n-1)\rho) & \text{with multiplicity }1 \text{ and eigenvector with constant weights,} \\[4mm]
\displaystyle \lambda_2=\sigma^2(1-\rho) & \text{with multiplicity } n-1 \; .
\end{array}
\right.
\end{equation}
Similarly, also the inverse matrix $\Sigma^{-1}$, can be computed $\forall n$.

\smallskip

As a first numerical example, let us consider a symmetric case $\mathbb{\mu} = \mu \, \textbf{1}$ as considered by \citet{Glasserman2014} in their numerical example.
In this case, the optimal portfolio in the nominal model and in the alternative model is the equally weighted one $\textbf{a}_0^\star=\textbf{1} / n $.
We also have an explicit expression for condition \eqref{eqn:conditiontheta} for the optimal portfolio in the alternative model $\textbf{a}_0^\star$.
Using  \eqref{eqn:eigenvaluesSigma}, it becomes 
\[
\theta<\theta_{max}(\textbf{a}_0^\star) = \dfrac{n}{\sigma^2\,(1+(n-1) \rho)} = \dfrac{n}{\lambda_1} \; .
\]


\smallskip

\begin{figure}
\centering
{\includegraphics[width=.49\textwidth]{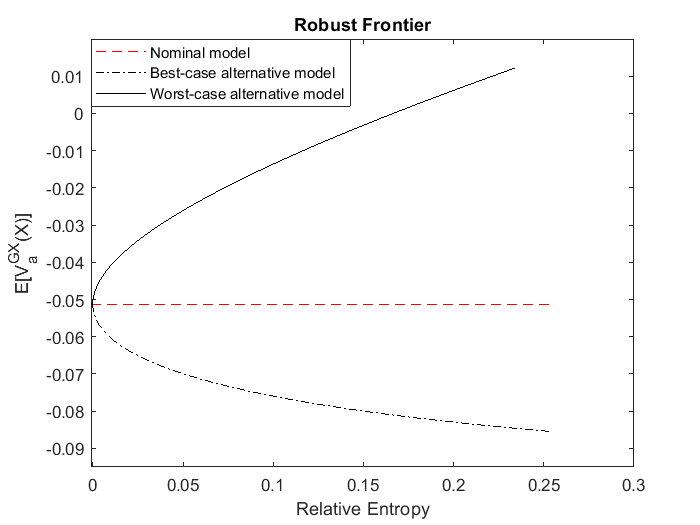}} 
{\includegraphics[width=.49\textwidth]{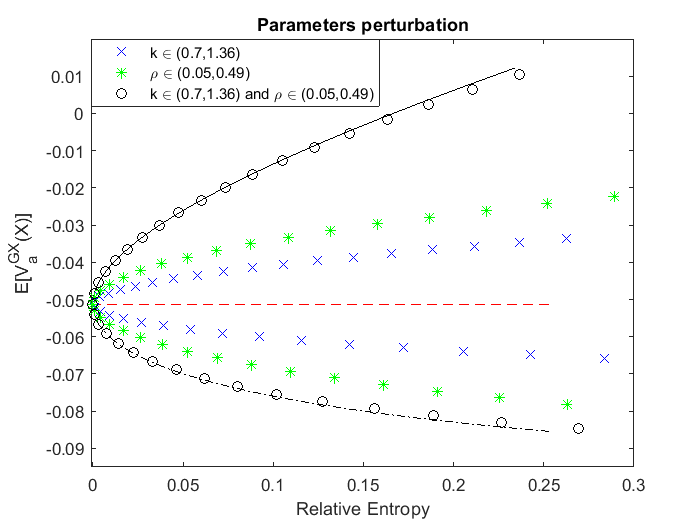}} 
\caption{\small
Value of the risk measure vs. relative entropy in the symmetric case. 
We consider 
$\gamma=1$,  $n=10$ assets,  
$\mu_i=0.1$, $\Sigma_{ii}=0.3$ $\forall i=1,\dots,10$ and $\rho=0.25$.
The left-hand figure shows the value of the risk measure, corresponding to the optimal portfolio, in the nominal (dashed red line) and in the alternative model (black line).
Considering the maximum relative entropy $\eta\in[0,0.25]$ as in \citet[][Figure 1, p.37]{Glasserman2014}, we compute, for each fixed value of $\eta$, 
the corresponding positive and negative values of $\theta^\star$ using \eqref{eqn:relentr_theta_GX}, with which we get the value of the risk measure \eqref{eqn:exp_utility_minimumvar}.
 In particular, negative $\theta^\star$ values correspond to the best-case alternative model (dotted-dashed line), 
while positive $\theta^\star$ values correspond to the worst-case approach (continuous line). 
The right-hand figure shows the value of the risk measure in the nominal model varying just the correlation parameter $\rho$ with $\rho\in[0.05, 0.45]$ (green stars) and 
a multiplicative parameter  $k$  of the variance parameter $\sigma^2$ with $k\in[0.72, 1.32]$ (blue crosses) as in \citet{Glasserman2014}. 
We notice that the value of the risk measure obtained varying both the parameters (black circles) is the same as that obtained in the alternative model (black line). Model risk in this case reduces simply to  {\it estimation} risk.}
\label{fig:frontieraeparametri}
\end{figure}

We consider exactly the same numerical example as that in \citet[][pp.36-37]{Glasserman2014} with $\gamma=1$,  $n=10$ assets,  
$\mu_i=0.1$, $\Sigma_{ii}=\sigma^2=0.3$ $\forall i=1,\dots,10$ and $\rho=0.25$ and we recall that they consider the special case with a mean-variance with fixed mean.
We plot the value of the risk measure for the optimal portfolio as a function of the maximum allowed relative entropy $\eta$ \citep[cf. Figure 1 in][p.37]{Glasserman2014}. 
The left-hand plot in Figure \ref{fig:frontieraeparametri} shows, for a set of maximum relative entropy values $\eta \in [0, 0.25]$,
the value of the risk measure in the alternative model in the worst-case approach (continuous black line) and in the best-case approach (dot-dashed black line).


\smallskip

As already stated, in this case,
due to Proposition \ref{prop:necsufcondition},
the optimal portfolios in the nominal and in the alternative model coincide. 
This is different from the result shown in \citet[][Figure 1, p.37]{Glasserman2014} 
in which 
these two optimal portfolios do not coincide.
This incoherence might be due to a slow convergence of the numerical algorithm. We return to this point in the following.

\smallskip

Moreover, 
\citet{Glasserman2014} claim that ``model error does not correspond to a straightforward error in parameters" (cf. p.37). 
To illustrate this idea --because in the mean-variance with fixed mean framework the alternative model differs from the nominal model just for the variance matrix-- these authors studied
the value of the risk measure obtained by varying just two parameters: the common correlation parameter $\rho$ and a parameter $k$ that multiplies 
the variance parameter $\sigma^2$ in the 
the variance matrix $\Sigma$. 
In particular, they let the correlation parameter vary
between $\rho=0.05$ and $\rho=0.45$, and the parameter $k$ vary between $k=0.72$ and $k=1.32$.

The result obtained by varying $k$ and $\rho$ separately is shown in \citet[][Figure 1, p.37]{Glasserman2014} and it is in agreement with blue crosses and green stars in the right-hand panel of Figure \ref{fig:frontieraeparametri}. 
The new result is that the perturbation of both parameters $\rho$ and $k$, in the same range as before, modifies the value of the risk measure that reaches the value obtained in the alternative measure (see black circles in Figure \ref{fig:frontieraeparametri}); i.e., in this framework model error can be completely explained as {\it estimation} error. 

\smallskip

\begin{figure}
\centering
{\includegraphics[width=.7\textwidth]{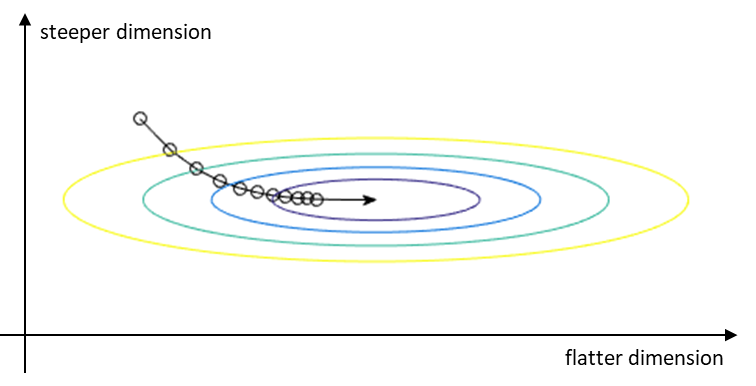}} 
\caption{\small
Evolution of a gradient descent algorithm in the minimisation of a quadratic function in two dimensions. Similar results hold for other first order optimisation algorithms. 
We notice that the algorithm is slow in reaching the correct optimum in the flatter dimension. In more dimensions, this behaviour is amplified and the algorithm could not reach the optimum with a reasonable precision.}
\label{fig:GradDesc}
\end{figure}

Furthermore, because we have in this case a complete analytical solution, we can understand the reason 
why a numerical approach 
can be slow.
Solving the optimisation problem \eqref{eqn:astartheta_minvar} is equivalent to selecting the minimum of a paraboloid.
A first order algorithm decreases faster in the direction of higher eigenvalues and more slowly when eigenvalues are lower.
For example, Figure \ref{fig:GradDesc} shows the evolution of the gradient descent numerical algorithm used in the minimisation of a quadratic function in two dimensions: 
the algorithm is fast in the direction with maximum variability but varies slowly in the direction of minimum variability; i.e., in the direction of the eigenvector of the matrix $\Sigma$ corresponding to the minimum eigenvalue. 
In our case, for every $\eta$ in the interval of interest, we have to solve an optimisation problem. 
Each optimisation has two main features, as shown in \eqref{eqn:eigenvaluesSigma}: i) the largest eigenvalue is almost $n$ times larger than the other ones and ii) there are $n-1$ minimum eigenvalues. 
Thus, the numerical algorithm becomes slow in the direction of minimum variability and it could stop before it reaches the correct optimum, in particular if $n$ is very large. 
This can be one reason why an analytical solution can be useful.

\bigskip

\begin{figure}
\centering
{\includegraphics[width=.7\textwidth]{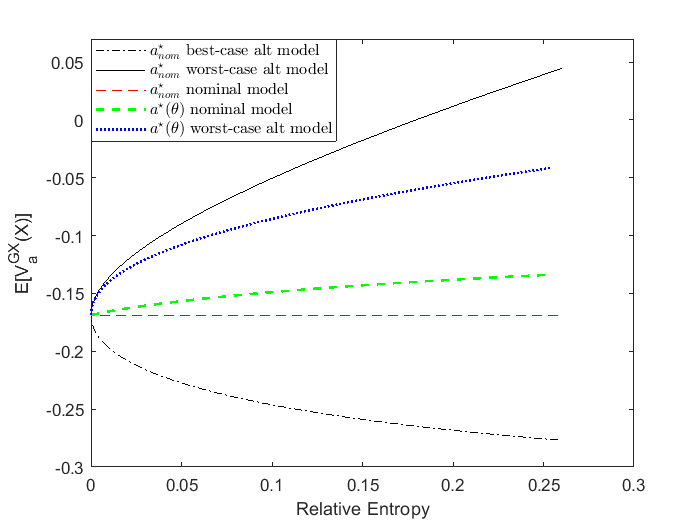}} 
\caption{\small
Value of the risk measure vs. relative entropy in a non-symmetric case.
We consider $\mu_i=0.1\cdot (1+x_i),\; i=1,\ldots,10$, with $x_i$ drawn from a standard normal distribution, while
the other parameters are the same of Figure 2. As in Figure \ref{fig:frontieraeparametri}, we have considered the relative entropy $\eta\in[0,0.25]$.
The dashed red line shows the value of the risk measure in the nominal model for the optimal nominal portfolio $a^\star_{nom}$ \eqref{eqn:astarclassic};
the black line is the value of the risk measure  in the alternative model in the worst-case approach (continuous line) and 
in the best-case approach (dotted-dashed line) for the optimal nominal portfolio $a^\star_{nom}$; 
the dashed green line shows the value of the risk measure in the nominal model for the robust portfolio $a^\star (\theta^\star)$ \eqref{eqn:system_astartheta_GX};
the dotted blue line shows the value of the risk measure in the alternative model for the robust portfolio. 
}
\label{fig:frontierNonSymm}
\end{figure}

Finally, as a second numerical example  we consider a non-symmetric case. 
The parameters are the same ones of previous numerical example, but with mean $\mu_i=0.1\cdot(1+x_i)$, $i=1,\ldots,10$, 
with $x_i$ drawn from a standard normal random distribution. The frontier obtained is shown in Figure \ref{fig:frontierNonSymm}.
In this case the robust portfolio,
i.e. the optimal portfolio in the worst-case alternative model  $a^\star (\theta^\star)$ \eqref{eqn:system_astartheta_GX}
where $\theta^\star$ is obtained via Proposition \ref{prop:thetastarsurface_GX}, 
 is different from the optimal portfolio in the nominal model $a^\star_{nom}$, 
as we have proven in Proposition \ref{prop:necsufcondition}. 
The result obtained in Figure \ref{fig:frontierNonSymm} looks similar to the one in \citet[][Figure 1, p.37]{Glasserman2014} and 
the two plots differ just for the values of the risk measure on the vertical axis, that depend on the chosen values for $\bm{\mu}$. 

In Figure \ref{fig:frontierNonSymm} we can observe the consequences of selecting the robust portfolio.  
On the one hand, the value of the risk measure  in the nominal model  is larger  
 for the robust portfolio (dashed green line) w.r.t. the one for the optimal nominal portfolio (dashed red line),
where the latter clearly does not depend on the relative entropy;  
on the other hand,  the value of the risk measure  in the alternative worst-case model (with variane $\tilde{\Sigma}$ in \eqref{eqn:mutilde_Sigmatilde}) is significantly lower
 for the robust portfolio  (dotted blue line)
 w.r.t. the one valued for the optimal nominal portfolio (continuous black line).

\section{Conclusions}

We have studied the effect of model risk on the optimal portfolio in the mean-variance selection problem. 
Model risk is measured via the worst-case approach, taking the relative entropy as measure of the divergence between the nominal and the alternative model;
in particular, we have considered all alternative models within a KL-ball $P_\eta$ of radius $\eta$.
When asset returns are modeled with a multivariate normal,
this problem has been  numerically solved for small $P_\eta$ by \citet{Glasserman2014} in the mean-variance case 
with an additional constraint on the mean, chosen 
equal to the one  in the nominal model. 

\bigskip
In this paper, 
we have analytically solved the optimal portfolio selection problem in the alternative model in a generic mean-variance framework for a generic KL-ball $P_\eta$.
We have proven that the optimal portfolio in the worst-case approach
 is unique and given by equation \eqref{eqn:system_astartheta}, 
where the optimal $\theta^{\star}$ is the unique positive solution of the equation $R(\theta^{\star}) = \eta$
with $R(\theta)$ given in \eqref{eqn:relentr}.

We have also solved the special case considered in
\citet{Glasserman2014}, in which they impose the alternative mean constrained 
(cf. solution \eqref{eqn:system_astartheta_GX}  
with entropy \eqref{eqn:relentr_theta_GX}).

Finally, we have analyzed in detail the situations when model risk and {\it estimation} risk coincide and 
we have shown two numerical examples.
In particular, one of these examples considers exactly the same illustrative example of \citet[][Figure 1, p.37]{Glasserman2014},  without been able to reproduce their numerical solutions.
This fact shows the relevance of the provided analytical solution in the worst-case approach for model risk
because in some cases solving the nested optimizations   (cf. problem \eqref{eqn:probLagrange_infinfsup}) can be a challenging operational research problem from a numerical point of view.


\section*{Acknowledgments}
The authors thank all participants to the seminar at the European Investment Bank (EIB)
and conference participants 
to the $19^{\text{th}}$ Quantitative Finance Workshop at ETH Zurich and 
to the SIAM Conference on Financial Mathematics \& Engineering $2019$ at University of Toronto. 
We are grateful in particular to Michele Azzone, Giuseppe Bonavolont\'a, Mohamed Boukerroui, Szabolcs Gaal, Juraj Hlinicky, Aykut Ozsoy, Oleg Reichmann, Sergio Scandizzo, Claudio Tebaldi, Pierre Tychon for useful comments.

The authors acknowledge EIB financial support under the EIB Institute Knowledge Programme.
The findings, interpretations and conclusions presented in this document are entirely those of the authors and should not be attributed in any manner to the EIB.
Any errors remain those of the authors. 

\bigskip

\bibliography{ModelRiskPortfolio}

\begin{thebibliography}{16}
\providecommand{\natexlab}[1]{#1}
\providecommand{\selectlanguage}[1]{\relax}

\bibitem[{Boyd and Vandenberghe(2004)}]{BoydConvexOptim}
Boyd, S. and Vandenberghe, L., 2004. \textit{Convex optimization}, Cambridge
  university press.

\bibitem[{Calafiore(2007)}]{Calafiore2007}
Calafiore, G.C., 2007. Ambiguous risk measures and optimal robust portfolios,
  \textit{SIAM Journal on Optimization}, 18~(3), 853--877.

\bibitem[{Gantmacher and Kre{\u\i}n(1960)}]{Gantmacher1960}
Gantmacher, F.R. and Kre{\u\i}n, M.G., 1960. \textit{Oszillationsmatrizen,
  oszillationskerne und kleine schwingungen mechanischer systeme}, vol.~5,
  Akademie-Verlag.

\bibitem[{Gilboa and Schmeidler(1989)}]{GilboaSchmeidler1989}
Gilboa, I. and Schmeidler, D., 1989. Maxmin expected utility with non-unique
  prior, \textit{Journal of Mathematical Economics}, 18~(2), 141--153.

\bibitem[{Glasserman and Xu(2014)}]{Glasserman2014}
Glasserman, P. and Xu, X., 2014. Robust risk measurement and model risk,
  \textit{Quantitative Finance}, 14~(1), 29--58.

\bibitem[{Hansen and Sargent(2008)}]{HansenSargent2008}
Hansen, L.P. and Sargent, T.J., 2008. \textit{Robustness}, Princeton university
  press.

\bibitem[{Harville(1997)}]{HarvilleMatrix}
Harville, D.A., 1997. \textit{Matrix algebra from a statistician's
  perspective}, vol.~1, Springer.

\bibitem[{Kerkhof \textit{et~al.}(2010)Kerkhof, Melenberg, and
  Schumacher}]{Kerkhof2010}
Kerkhof, J., Melenberg, B., and Schumacher, H., 2010. Model risk and capital
  reserves, \textit{Journal of Banking \& Finance}, 34~(1), 267--279.

\bibitem[{Kullback and Leibler(1951)}]{KullbackLeibler}
Kullback, S. and Leibler, R.A., 1951. On information and sufficiency,
  \textit{The annals of mathematical statistics}, 22~(1), 79--86.

\bibitem[{Lam(2016)}]{Lam2016}
Lam, H., 2016. Robust sensitivity analysis for stochastic systems,
  \textit{Mathematics of Operations Research}, 41~(4), 1248--1275.

\bibitem[{Li and Ng(2000)}]{LiNg}
Li, D. and Ng, W.L., 2000. Optimal dynamic portfolio selection: Multiperiod
  mean-variance formulation, \textit{Mathematical {F}inance}, 10~(3), 387--406.

\bibitem[{Markowitz(1952)}]{Markowitz1952}
Markowitz, H., 1952. Portfolio selection, \textit{The Journal of Finance},
  7~(1), 77--91.

\bibitem[{Merton(1972)}]{Merton1972}
Merton, R.C., 1972. An analytic derivation of the efficient portfolio frontier,
  \textit{Journal of Financial and Quantitative Analysis}, 7~(4), 1851--1872.

\bibitem[{Penev \textit{et~al.}(2019)Penev, Shevchenko, and Wu}]{Penev2019}
Penev, S., Shevchenko, P.V., and Wu, W., 2019. The impact of model risk on
  dynamic portfolio selection under multi-period mean-standard-deviation
  criterion, \textit{European Journal of Operational Research}, 273, 772--784.

\bibitem[{Protter and Morrey(2012)}]{Protter}
Protter, M.H. and Morrey, C.B.J., 2012. \textit{Intermediate calculus},
  Springer Science.

\bibitem[{Sherman and Morrison(1950)}]{ShermanMorrison}
Sherman, J. and Morrison, W.J., 1950. Adjustment of an inverse matrix
  corresponding to a change in one element of a given matrix, \textit{The
  Annals of Mathematical Statistics}, 21~(1), 124--127.

\end{thebibliography}
\bibliographystyle{tandfx}

\section*{Shorthands and notation}

{\bf Shorthands}

\[
\begin{array}{lcl}
A,\,B,\,C,\,D & : & \text{defined in \eqref{eqn:ABCD}} \\[1mm]
\text{cf.} & : & \text{compare; from Latin: confer}\\[1mm]
\text{e.g.} & : & \text{for example; from Latin: exempli gratia}\\[1mm]
\text{i.e.} & : & \text{that is; from Latin: id est}\\[1mm]
\text{KL-ball} & : & \text{Ball of } m \text{ where the Kullback-Leibler entropy is lower than } \eta \\[1mm]
\text{p.d.f.} & : & \text{probability density function} \\[1mm]
\text{r.v.} & : & \text{random variable}\\[1mm]
\text{s.t.} & : & \text{such that}   \\[1mm]
{\rm w.r.t.} & : & {\rm with \; respect \;  to} 
\end{array} 
\; 
\]

\newpage

{\bf Notation}

\begin{center}
\begin{tabular}{|l|l|} \hline
{\bf Symbol} & {\bf Description} \\
\hline
$\textbf{1}$ & Vector of all $1$s in $\mathbb{R}^n$ \\[1mm]
$\alpha$ & Lagrange multiplier \\[1mm]
$\textbf{a}$ & Portfolio's weights vector, named portfolio \\[1mm]
$\textbf{a}_{\text{nom}}^\star$, $\textbf{a}^\star(\theta)$ & Optimal portfolio in the nominal measure and in the alternative measure \\[1mm]
$\mathcal{A}$ & Domain for the portfolio $\textbf{a}$, satisfying the constraint $\textbf{1}^T \textbf{a} = 1$ \\[1mm]
$\gamma$ & Risk aversion parameter \\[1mm]
$\Gamma(S;\theta,\gamma)$ & Defined in \eqref{eq:Gamma} \\[1mm]
$\Gamma^{GX}(S;\theta,\gamma)$ & Defined in \eqref{eqn:GammaGX} \\[1mm]
$d$, $d^\star$ & Dual Lagrangian problem corresponding to $p$ and related optimal value \\[1mm]
$\mathcal{D}$ & Domain for $(\theta,\textbf{a})$ defined in \eqref{eqn:D} \\[1mm]
$f({\bf{X}}), \tilde{f}({\bf{X}})$ & Probability densities in nominal and alternative model \\[1mm]
$\eta$ & Maximum KL-divergence between the alternative and the nominal one \\[1mm]
$\theta$, $\theta^\star$ & Optimisation parameter and corresponding optimum \\[1mm]
$\theta_{\text{max}}(\textbf{a})$ & Upper bound for $\theta$ defined in \eqref{eqn:conditiontheta} \\[1mm]
$\tilde{\theta}$ & Argmin in $\theta$ of $\mathcal{L}(\theta,\textbf{a})$ for a given portfolio $\textbf{a}$ \\[1mm]
$I$ & Identity matrix \\[1mm]
$k$ & Multiplicative parameter of the variance parameter $\sigma^2$ \\[1mm]
$\lambda_i$, $\tilde{\lambda}_i$ & Eigenvalues of the variance matrices $\Sigma$ and $\tilde{\Sigma}$ \\[1mm]
$\mathcal{L}(\theta,\textbf{a},m(\textbf{X}))$ & Lagrangian function associated to constrained maximisation problem \eqref{eqn:probopt1infsup} \\[1mm]
$\mathcal{L}(\theta,\textbf{a})$ & Lagrangian function computed in the optimal change of measure \eqref{eqn:mstarthetaa} \\[1mm]
$m({\bf{X}})$ & Change of measure, defined as $\tilde{f}({\bf{X}})/{f(\bf{X})}$ \\[1mm]
$m^\star_{\theta, \textbf{a}} ({\bf{X}})$ & Worst-case change of measure depending on parameter $\textbf{a}$ \\[1mm]
$m^\star_{\theta} ({\bf{X}})$ & Worst-case change of measure corresponding to the optimal portfolio $\textbf{a}^\star(\theta)$ \\[1mm]
$m^{\star p}({\bf{X}})$, $m^{\star d}_{\theta^\star}({\bf{X}})$ & Optimal change of measures for the primal and dual problem $p$ and $d$ \\[1mm]
$\bm{\mu}$, $\tilde{\bm{\mu}}$ & Mean vector in the nominal and in the alternative model \\[1mm]
$n$ & Number of assets considered \\[1mm]
$p$, $p^\star$ & Primal problem optimisation in \eqref{eqn:probopt1infsup} and related optimal value \\[1mm]
$P_\eta$ & KL-ball with radius $\eta$ \\[1mm]
$\rho$ & Correlation parameter \\[1mm]
$R(\tilde{f},f)$ & Relative entropy function between  nominal and alternative models \\[1mm]
$R(\theta,\textbf{a})$ & Relative entropy corresponding to the optimal change of measure \eqref{eqn:mstarthetaa} \\[1mm]
$R(\theta)$ & Relative entropy corresponding to the optimal portfolio $\textbf{a}^\star(\theta)$ \\[1mm]
$S$, $S^\star(\theta)$ & Defined as $\textbf{a}^T\Sigma\textbf{a}$ for a portfolio $\textbf{a}$ and for the optimal one $\textbf{a}^\star(\theta)$ \\[1mm]
$\Sigma$, $\tilde{\Sigma}$ & Variance matrix in the nominal and in the alternative model \\[1mm]
$\sigma^2$ & Variance parameter  \\[1mm]
$V_\textbf{a}({\bf{X}})$ & Measure of risk associated with $\bf{X}$ and with parameter $\textbf{a}$ \\[1mm]
$V^{GX}_\textbf{a}({\bf{X}})$ & Measure of risk in the special case with constant mean \\[1mm]
$\bf{X}$ & Stochastic asset returns \\[1mm]
\hline
\end{tabular}
\end{center}

\newpage

\section*{Appendix A}

In this appendix
we prove Lemma \ref{lem:relentr_monotone} on the monotonicity of the relative entropy function.
We also state and prove a technical Lemma, which is useful to compute the eigenvalues of the nominal variance-covariance matrix $\Sigma$ in the fully symmetric case.

\bigskip

{\bf Proof of Lemma \ref{lem:relentr_monotone}}.
For any value of $\textbf{a}$ and $\theta$ s.t. $\exists$ a solution $ m^\star_{\theta,\textbf{a}}(\textbf{X})$ \eqref{eqn:mstarthetaa}, we get
\begin{equation*}
R(\theta,\textbf{a})=\mathbb{E}\left[ m^\star_{\theta,\textbf{a}}(\textbf{X}) \ln{m^\star_{\theta,\textbf{a}}(\textbf{X})} \right]=\theta\mathbb{E}\left[ m^\star_{\theta,\textbf{a}}(\textbf{X}) V_\textbf{a}(\textbf{X}) \right] -\ln{\mathbb{E}[\exp(\theta V_\textbf{a}(\textbf{X}))]} \, .
\end{equation*}
We now study the behaviour of this function. 
First, we notice that the function is continuous 
$\forall \theta$ s.t. $ m^\star_{\theta,\textbf{a}}(\textbf{X})$ is well-defined. 
Then, to evaluate the slope of the relative entropy, we compute the first derivative.
Supposing to choose $ m_{\theta,\textbf{a}}(\textbf{X}) $ 
within a class of sufficiently regular functions, we can exchange the derivative with the expected value \citep[cf. e.g., ][Th.4, p.429]{Protter}. 
We get
\begin{align}
\frac{\partial R(\theta,\textbf{a})}{\partial \theta} &= \mathbb{E}\left[ m^\star_{\theta,\textbf{a}}(\textbf{X}) V_\textbf{a}(\textbf{X}) \right] + \theta \frac{\partial}{\partial \theta}\mathbb{E}\left[ m^\star_{\theta,\textbf{a}}(\textbf{X}) V_\textbf{a}(\textbf{X}) \right] - \frac{1}{\mathbb{E}\left[ \exp(\theta V_\textbf{a}(\textbf{X}) ) \right]} \frac{\partial}{\partial \theta} \mathbb{E}\left[ \exp(\theta V_\textbf{a}(\textbf{X})) \right]= \notag \\
&= \theta \left( \mathbb{E}\left[ m^\star_{\theta,\textbf{a}}(\textbf{X}) V_\textbf{a}^2(\textbf{X}) \right] - \mathbb{E}\left[ m^\star_{\theta,\textbf{a}}(\textbf{X}) V_\textbf{a}(\textbf{X}) \right]^2 \right) = \notag \\ 
&= \theta \, \text{Var}_{\tilde{f}}(V_\textbf{a}(\textbf{X})) \label{eqn:deR_detheta} \, , 
\end{align}
where $\text{Var}_{\tilde{f}}(V_\textbf{a}(\textbf{X}))$ is the variance of $V_\textbf{a}(\textbf{X})$ in the alternative measure.

Being the variance non-negative, the sign of the derivative, i.e. the slope of the relative entropy function, depends only on the sign of $\theta$ and 
the relative entropy is a monotone increasing function for positive values of $\theta$, and a monotone decreasing function for negative $\theta$
$ \qquad\hspace*{\fill}  \clubsuit \; $

\bigskip

\begin{lemma}
\label{lem:eigenSigma}
Given a matrix $K\in\mathbb{R}^{n\times n}$ with two values, $c$ on the diagonal and $d$ extra-diagonal, i.e. of the form
\begin{equation*}
K=\begin{bmatrix}
c & d & \dots & d \\
d & c & \dots & d \\
\vdots & \vdots & \ddots & \vdots \\
d & d & \dots & c
\end{bmatrix} \; ,
\end{equation*}
$K$ has an eigenvalue $\lambda_1=d\,(n-1)+c$ with eigenvector with constant weights and the remaining $n-1$ eigenvalues $\lambda_2=\dots=\lambda_n=c-d$.
\end{lemma}

\bigskip
{\bf Proof}.
We write $K$ as 
\begin{equation*}
K=d\,\textbf{1}\textbf{1}^T+(c-d)\,I \; .
\end{equation*} 
We simply have to find eigenvalues of $\textbf{1}\textbf{1}^T$ because the eigenvalues of $K$ are the sum of eigenvalues of $d\,\textbf{1}\textbf{1}^T$ and $(c-d)\,I$. 
The matrix $\textbf{1}\textbf{1}^T$ has one eigenvalue equal to $n$ corresponding to the eigenvector $\textbf{1} / {n} $. The remaining $n-1$ eigenvalues are all $0$ with eigenvectors equal to any basis of the kernel (due to the rank-nullity theorem).
We then get the eigenvalues and the eigenvectors of $K$ $ \qquad\hspace*{\fill}  \clubsuit \;$

\end{document}